\documentclass{eas}
\usepackage{graphicx}
\TitreGlobal{Physics and Astrophysics of Planetary Systems}
\begin{document}

\title{The Atmospheres of Extrasolar Planets} 
\author{Mark S.\ Marley}\address{Mail Stop 245-3; NASA Ames Research Center; Moffett Field, California 94035; USA}

\begin{abstract}
The characteristics of irradiated solar system planetary atmospheres have been studied for decades, consequently modern planetary science benefits from an exhaustive body of ground- and space-based data.  The study of extrasolar planetary atmospheres, by contrast, is still in its infancy and currently rests on a few score of datapoints, mostly of the transiting planets.  This short survey aims not to review this dynamic field but rather stresses the importance of a few theoretical concepts and processes for our understanding of exoplanet atmospheres.  Topics covered include atmospheric structure and dynamics, cloud processes and photochemistry of planetary atmospheres. Influences on the albedos, spectra, and colors of extrasolar planets are reviewed and caution is urged in the interpretation of exoplanet colors.  

\end{abstract}
\maketitle
\section{Introduction}
The atmosphere controls a planet's evolution through time and provides a window into the chemical and physical conditions under which the planet formed.  Except for airless worlds, like Mercury, the atmosphere also mediates the flow of information we receive about the nature of the planet.  In this contribution I aim to discuss a sampling of a few of the more important concepts which relate to our understanding of extrasolar planetary atmospheres.   I will focus on topics that are important for understanding spectral and photometric data from extrasolar planets, with an eye towards providing examples from the solar system.

In some sense the trajectory of extrasolar planet science is recapitulating the history of solar system planetary science.   Three decades ago in the University of Arizona Space Science series book ``Jupiter'' Wallace (1976) reviewed the sparse data then available on Jupiter's thermal emission spectrum and presented models of the planet's atmospheric structure.  His effort to piece together a consistent picture of the atmosphere from a handful of thermal infrared photometric measurements, reflection spectra, and transmission data from stellar occultations bears a striking resemblance to current efforts to understand the transiting extrasolar planets.  

Today, with our eyes now fixed on exoplanets, we are fortunate to have at hand the insights gained from such work over the past half century of planetary exploration. For a few transiting planets we again find ourselves in the situation where we have a few broad band brightness temperature measurements from which we are attempting to infer global atmospheric processes.  As we carry out our science we would do well to remember that this ground has been crossed before: irradiated planetary atmospheres have been encountered before the $21^{\rm st}$ century (a fact sometimes overlooked in the modern literature).  The excitement of our time, of course, is that the range of parameter space, from planetary masses to the dizzying range of incident radiation is far larger than previously encountered.  Nevertheless we should not be surprised that well understood processes familiar from solar system planetary science, including photochemistry, hot stratospheres, and cloud processes also play important roles in extrasolar planet atmospheres (Marley 1997).

To aid those attempting to traverse this new landscape of theory and data of exoplanet atmospheres for the first time,  I discuss in this chapter a few elementary themes that are helpful for understanding the conceptual scenery.  
While I will illustrate concepts by drawing from work in the field, I am not in any sense attempting to provide a comprehensive review of exoplanet science, as many such recent reviews are available (e.g., Charbonneau (2008), Deming (2008), and Seager et al.\ (2008) review recent progress; Burrows et al. (2001, 2006) and Barman et al. (2005) present useful surveys of much of the theory of these objects; and Marley et al.\ (2007) review the pre-2007 field).  Instead I present a brief, illustrated (if somewhat idiosyncratic) guidebook to a few of the important processes encountered at these new worlds.  Hence I will touch on atmospheric structure, cloud formation, the interpretation of brightness temperature and the importance of photochemistry. I will mention some of the processes contributing to stratospheric heating and conclude with some brief comments about albedos and colors of planets.

\section{Atmospheric Structure}

The basic relationships governing a static atmosphere in hydrostatic equilibrium are well known and presented in introductory textbooks, such as Chamberlain \& Hunten (1987).  Such texts make for excellent introductory reading as many of the fundamental relationships and processes governing extrasolar planets are of course familiar from the solar system planets.  One particularly useful relation can be derived for an ideal gas atmosphere in hydrostatic equilibrium. The number density of molecules per unit area in a gaseous column above a specified height, $\cal N$, can be related to the local number density of molecules, $n$, by
$${\cal N}\approx {p(z)\over{g(z) m}} = n(z) H. \eqno(1)$$
Here  the mean molecular weight is $m$, altitude is $z$ and $p,\,g$ and $H$ are the pressure, gravity and the scale height; the equality is exact for an isothermal atmosphere.

\begin{figure}
\includegraphics[angle=0,scale=0.5]{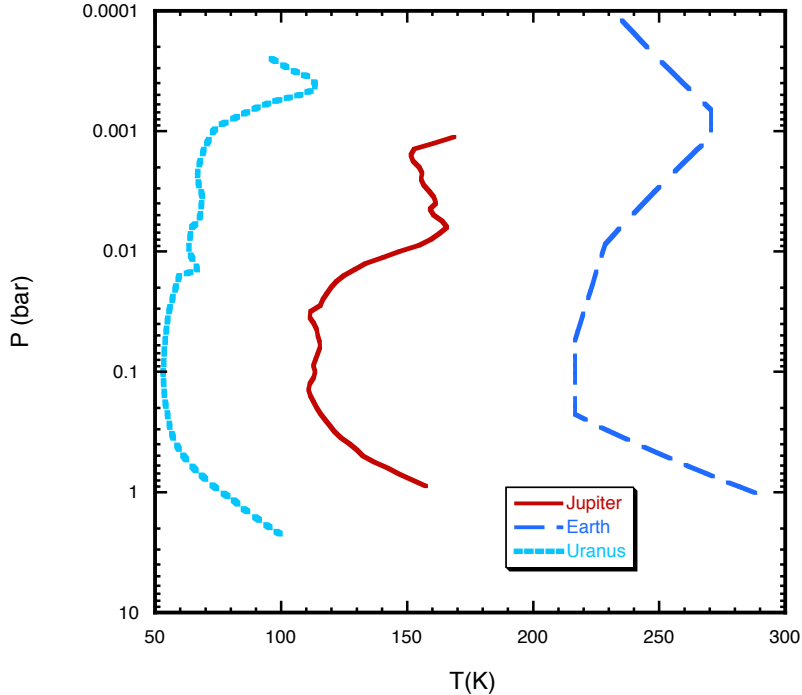}
\caption{Atmospheric temperature-pressure profiles for Uranus, Jupiter, and Earth.  In all three planets the temperature increases with depth below a few hundred millibars.  In each planet this atmospheric region--the troposphere--transports heat by convection from the deep interior, in the case of the giants, or the surface in the case of Earth. In each atmosphere the temperature  also rises at low pressure--the stratosphere--owing to the absorption of a fraction of the incident ultraviolet light by photochemical products (ozone in the case of Earth, various hydrocarbon products in the giants' atmospheres). Uranus and Jupiter data are from the Voyager Radio Science occultation experiments and the Earth profile is from the 1976 Standard Atmosphere.  All data are available on-line
(http://atmos.nmsu.edu/planetary\_datasets/indextemppres.html).}
\end{figure}

Equation (1) is useful for understanding a variety of basic characteristics of atmospheres.  For example if a molecular absorber has a cross section for interaction with radiation of $\sigma\,$($\rm cm^2$), then the column optical depth above a given pressure surface is $\tau_{\rm col} \sim {\cal N} \sigma$. All else being equal, $\tau_{\rm col} $ falls linearly with pressure.
In a typical planetary atmosphere energy is transported upwards by convection until the atmosphere becomes optically thin to thermal radiation, when $\tau_{\rm col}\sim 1$.  Above this level the outgoing energy is transported by radiation.  The pressure level in the atmosphere of this radiative-convective boundary clearly depends upon the composition of the atmosphere, the opacity of the major atmospheric constituents, gravity and the temperature.  For example, in the Earth's atmosphere the surface temperature is about $290\, \rm K$.  Only at a temperature of about $220\,\rm K$ near $100\,\rm mb$ (at the tropopause) is the optical depth at the peak of the Planck function low enough that the air can radiate efficiently to space (Figure 1).  At lower pressures still a gray atmosphere would reach a constant temperature, known as the skin temperature $T_0 = (1/2)^{1/4} T_{\rm eff}$, where  $T_{\rm eff}$ is the effective temperature.

\begin{figure}
\includegraphics[angle=0,scale=0.7]{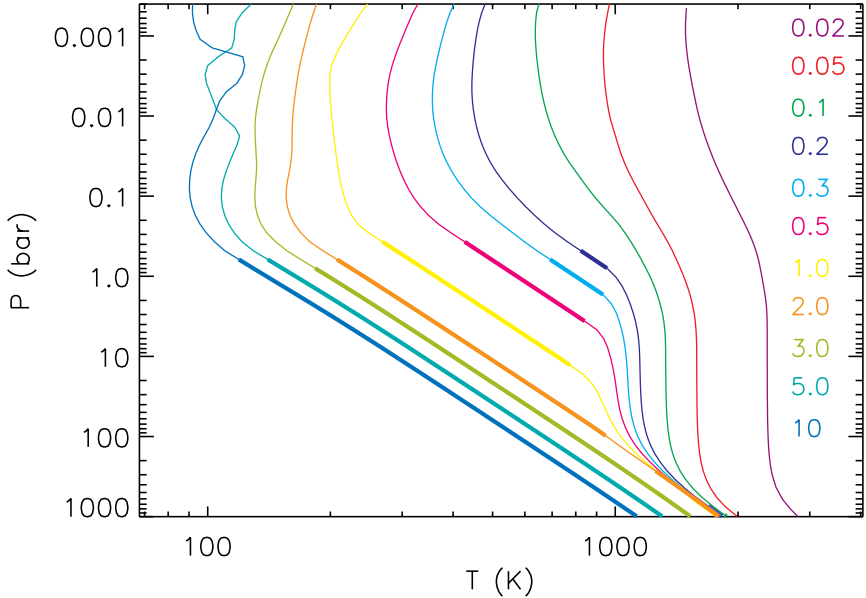}
\caption{Pressure-temperature profiles for $\sim 4.5\,\rm Gyr$-old Jupiter-like planets  from 0.02
to 10 AU (left to right) from a solar-type star.  Thick lines
are convective regions while thin lines are radiative regions. The profiles at 5 and 10 AU show deviations that
arise from numerical noise in the chemical equilibrium table near condensation points, but this has a negligible
effect on planetary evolution.  Figure adapted from Fortney et al.\ (2007).}
\end{figure}

The column abundance of molecules above a given pressure level in the atmosphere controls the level at which the atmosphere becomes optically thin to outgoing
thermal radiation, and hence the location of the tropopause.  The tropopause pressure will thus vary with gravity as well as with atmospheric structure and composition.
In a lower gravity atmosphere each molecule ``weighs'' less, so the column number of molecules, $\cal N$ above a given pressure level must be larger to compress the gas than in a higher gravity atmosphere.  Indeed Equation (1) tells us that the column optical depth is inversely proportional to $g$, so the column number density above the 1 bar surface on a planet with half Earth's gravity and a similar atmosphere would be twice that of our atmosphere.   All else being equal, the tropopause would be at a lower pressure.  For gas giant planets, which tend to have roughly constant radii regardless of mass, higher mass generally corresponds to higher gravity and thus more transparent atmospheres to a given pressure level.  Given the actual complexity of varying depths of absorption of incident radiation with variations in gravity and atmospheric composition, theoretical models, which relate atmospheric temperature to pressure, are required to fully recognize the subtleties of atmospheric structure.  Such models are shown in Figure 2.

\section{Atmospheric Dynamics}

The overall structure of an irradiated atmosphere depends both upon the depth at which incident energy is absorbed and internal sources of energy.  For a giant planet, convection transports energy outwards from the planetary interior, as the planet cools slowly over time.   For planets more distant than a few AU from their primary star most incident energy is absorbed fairly deep in the atmosphere, below the depth at which the atmosphere becomes optically thick in the thermal infrared (Fortney et al. 2007). This is because most gasses are more transparent in the optical than in the infrared.  As a result the absorbed incident energy simply adds to the internal energy being transported outwards by convection and the temperature profile resembles that of Jupiter shown in Figure 1.  Another consequence of this relatively deep deposition of solar radiation is that the internal heatflow is preferentially transported by convection to the poles of Jupiter yielding a relatively isothermal planet at the radiative-convective boundary (Ingersoll \& Porco 1978).

However, for those giant planets found closer to their primary stars, the radiative-convective boundary is deeper but absorption of incident flux still occurs at a similar altitude (to the extent that composition is unchanged). Thus the large incident flux upon a hot Jupiter is absorbed above the radiative-convective boundary (Figure 2).  As a consequence an isothermal layer appears between the top of the deep convective zone and the region of the atmosphere in which incident flux is absorbed.  In this case the deep internal heatflow is distinct from the thermalized incident radiation and the global temperature distribution is no longer relatively homogeneous and equator to pole temperature gradients can be large.  

The nature of the atmospheric circulation of giant planets, which is ultimately driven by the various energy fluxes, depends as well on a variety of influences, particularly including rotation rate and atmospheric scale height.  Showman et al.\ (2007) present a very useful introduction and review of this topic.  The atmospheric redistribution of energy by winds is unquestionably of paramount importance for the hot Jupiters and the efficiency of redistribution controls the global temperature map and consequently the phase variation of thermal emission, which has been successfully measured for multiple planets by {\it Spitzer} (notably by Knutson et al. 2007; Cowan \& Agol give a complete summary of observations through mid-2008).  Since a planet's thermal emission can arise from different depths in the atmosphere at different wavelengths (\S5), any variation in redistribution efficiency with altitude will manifest itself as differing thermal emission maps  as a function of wavelength (Burrows et al.\ 2008a; Fortney et al.\ 2008).  Ultimately coupled models of radiative transfer and dynamics, similar to terrestrial global circulation models, will be required to understand all of the contributing factors (Showman et al.\ 2008 present one such model and review recent progress in the field).

\section{Clouds}

Clouds play an important role in planetary atmospheres.  They scatter incident light back to space and sequester condensed species from the overlying atmosphere.  Jupiter's
atmosphere provides a point of departure for understanding the diversity of giant planet atmospheres that we expect to encounter outside of the solar system (see Lodders 2006 for a full discussion).

\begin{figure}
\includegraphics[angle=0,scale=1]{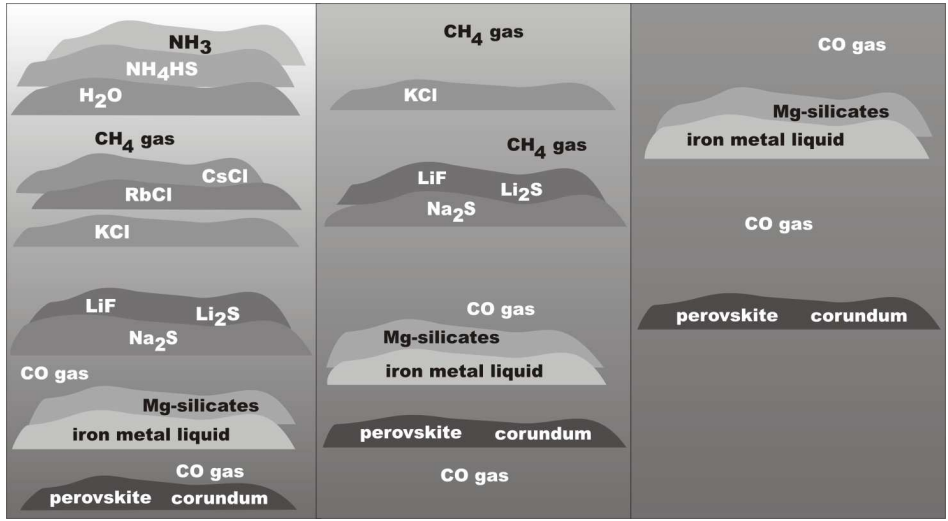}
\caption{Cloud structure expected to be found in, from left to right, Jupiter's atmosphere, a Jupiter-like planet too warm for water clouds to form, and a hot Jupiter.  Note that the overall cloud structure is similar in each panel, but cloud levels move upward as the atmosphere is warmed.  Important labeled condensates include perovskite ($\rm CaTiO_3$) which removes TiO from the gas phase, the salts $\rm Na_2S$ and KCl which remove two important alkali absorbers (Figure 7) from the gas, and water clouds which appear near the top of the Jupiter-like structure on the left.   Figure modified from Lodders (2004).}
\end{figure}

To understand Jupiter's cloud structure we might imagine an air parcel moving upwards from the deep interior (Figure 3).  We start at a temperature of about $2000\rm\, K$ and slowly raise the gas parcel up; as the gas rises, it cools adiabatically. In the deep atmosphere all gasses are well mixed.  At the point in the atmosphere where a condensible species' 
saturation mixing ratio first equals the vapor mixing ratio, a cloud base is found.   The first constituents to condense are refractory oxides such as perovskite and corundum, followed by various magnesium silicates including enstatite and forsterite.  Iron is predicted to condense as a native metal.  Although we cannot see this region of Jupiter's atmosphere, we know that iron clouds are there because hydrogen sulfide gas is detected in Jupiter's visible atmosphere\footnote{If iron grains were distributed uniformly above the iron condensation layer,  $\rm H_2S$  gas would react to form iron sulfide, FeS, thus removing sulphur-bearing gasses from the atmosphere (Lodders 2004).} (Niemann et al. 1998).  As we move upwards in the atmosphere the temperature continues to fall and eventually water clouds form, removing $\rm H_2O$ from the gas phase.  Above the water clouds, the atmosphere continues to cool until ammonia clouds form.  It is the ammonia clouds of Jupiter, dusted by various photochemical pollutants, that we see reflecting sunlight back from the planet.  In \S9 we'll consider how giant planets somewhat different than Jupiter might appear.

Since clouds scatter and absorb incident stellar radiation as well as emergent thermal radiation, they play a very important role in controlling the appearance and thermal structure of a planet.  However clouds are intrinsically difficult to model from {\it a priori} physical considerations.  The detailed behavior of terrestrial cloud cover as a function of atmospheric temperature is the leading source of uncertainty in global climate models, for example.  Although the chemistry is thought to be well understood, predicting cloud behavior for extrasolar planets, including such issues as particle sizes, vertical distribution, and any horizontal patchiness is difficult. Nevertheless some efforts at cloud modeling have been made (see the recent review by Helling et al. (2008)).  The hot L-type ultracool dwarfs have atmospheres with thick silicate and iron clouds. Accounting for the effects of these clouds has proven challenging.  Given the difficulty of cloud modeling in general and our experience with brown dwarfs, it seems that model predictions for the  spectra of extrasolar planets must be regarded skeptically, at least when clouds are expected.

\section{Brightness Temperature}

The spectrum of any planet is composed of two components: scattered incident radiation from the planet's star and thermal emitted flux from the planet.  The thermal flux represents both energy arising from processes interior to the planet and re-radiated absorbed incident radiation.  For solar system planets these two components of the spectrum are usually well separated in wavelength, but for the hottest exoplanets there can be substantial overlap between thermal radiation and scattered incident light.  
For a planet, such as a transiting planet, with a known radius, the thermal emission spectrum is often equated for convenience, wavelength by wavelength, to the thermal emission from a blackbody.  For an isothermal solid sphere with emissivity unity the observed spectrum would equal that of a blackbody with a fixed temperature.  However for a real planet the flux will differ from that of a blackbody with the same radius and at each wavelength a ``brightness'' temperature $T_B(\lambda)$ may be defined.  

\begin{figure}
\includegraphics[angle=0,scale=0.6]{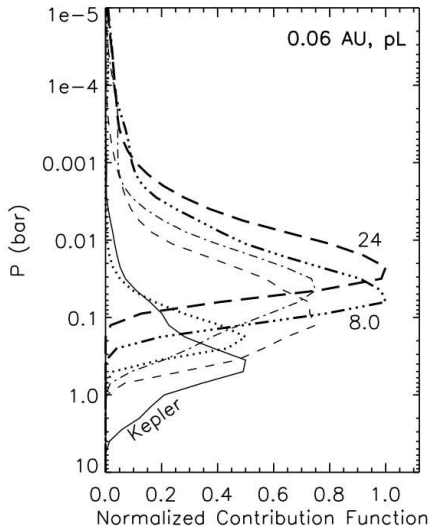}
\caption{Contribution functions calculated for a cloudless hot-Jupiter lacking a stratosphere. 
Contribution functions are calculated for various Spitzer broadband filters (the unlabeled curves correspond to various IRAC filters), K band, and the Kepler band at 450--900 nm (black solid curve). For clarity some of the curves have been normalized to 0.5 or 0 .75 rather than 
1. Figure from Showman et al. (2008).}
\end{figure}

Although thermal emission data for transiting planets are often reported in terms of of $T_B(\lambda)$, such data must be regarded with some care.  Except in special cases (e.g., an isothermal atmosphere) brightness temperature is not a measure of physical temperature or effective temperature.  Rather it gives a weighted measure of atmospheric temperatures over a range of pressures from which flux emerges from the planet.  To see this it is useful to consider the expression for the upwards or outgoing
intensity measured from an atmosphere as a function of frequency, $I_\nu(0, \mu)$, where $\mu$ is the cosine of the angle from the vertical.  Making use of Eq. (1):
$$I_\nu(0, \mu) = {{\sigma_\nu n(z_0)}\over \mu } \int\limits_0^\infty B_\nu(T) \exp {\biggl (}-{{z-z_0}\over H}-{\tau_\nu \over \mu} {\biggr )}dz. \eqno(2) $$
The exponential function in this equation, known as a weighting function (see Chamberlain \& Hunten 1987), describes the relative contribution to the outgoing flux from each  altitude $z$, defined here as the distance from the peak of the function, $z_0$.  Crudely the brightness temperature can indeed be equated to the temperature at the peak of the contribution function with height.  More importantly,  the emergent flux, expressed as a single brightness temperature is actually measuring thermal emission from a range of altitudes in the atmosphere around $z_0$ with varying physical temperatures.
Since $\tau_\nu$ can vary dramatically with frequency $\nu$, the contribution function---and thus the brightness temperature---can be quite different at different wavelengths.
For example Figure 4 illustrates theoretical contribution functions for several different commonly used {\it Spitzer} filters applied to the atmosphere of a cloud-free hot Jupiter (Showman et al.\ 2008).  Note that for each bandpass the measured emitted flux 
emerges over different vertical regions of the atmosphere with different temperatures. In regions of low opacity one sees deeper into the atmosphere which, for a monotonically increasing temperature profile with depth, means higher brightness temperature.  High opacity spectral regions correspond to lower pressures and lower temperatures.  However if there is an inverted temperature profile then a situation could emerge where $T_B(\lambda_1) > T_B(\lambda_2)$ and $T_B(\lambda_2) < T_B(\lambda_3)$ where $\tau(\lambda_1) > \tau(\lambda_2) > \tau(\lambda_3)$.  This is a commonplace occurrence in the atmospheres of solar system giants and the chromospheres of stars.

Furthermore, because of limb effects, this range in pressures from which flux emerges also varies over the disk.  Thus even for a gray atmosphere, with constant optical depth as a function of wavelength, the brightness temperature is in general not equal to the effective temperature at all wavelengths.

For these reasons, while brightness temperatures are useful shorthands to convey information about planetary spectra, they must be regarded with some caution.  For example $8\,\rm \mu m$ Spitzer observation of 
 the hot Neptune GJ 436b
(Deming et al. 2007, Demory et al. 2007) yield a brightness temperature  of $712 \pm 36\,\rm K$
which is modestly above the predicted effective temperature (Deming et al. 2007).  Since we do not expect, in general for $T_B = T_{\rm eff}$ the information content of this single datapoint is limited.  With atmosphere models and additional data points the value of each brightness temperature measurement increases, as was the case with Wallace's 1976 study of Jupiter cited in the introduction.

\section{Photochemistry}
  \begin{figure}
\includegraphics[angle=0,scale=0.5]{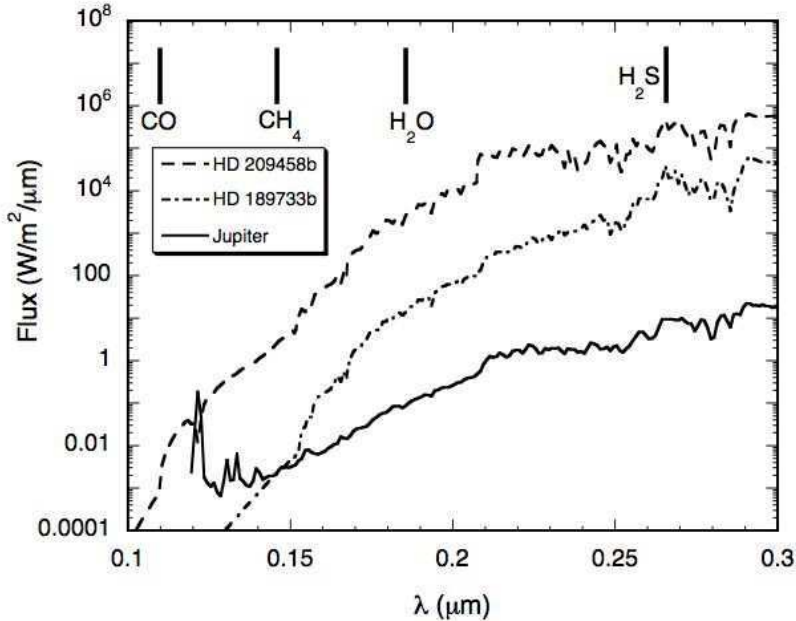}
\caption{Incident flux at the top of the atmospheres of two transiting planets compared to that received by Jupiter. Vertical lines 
denote the approximate maximum wavelengths at  which various molecules can be dissociated (after Marley et al. 2007). Given the large incident fluxes received by these and other hot Jupiters, many of the most abundant atmospheric species will be easily photolyzed, which will likely lead to a rich photochemistry. }
\end{figure}
Atmospheric molecules can be dissociated by the absorption of ultraviolet light, a process that happens high in the atmosphere before most incident UV light is scattered back to space.  Photochemical products can then participate in complex reaction chains, producing various molecular products.  A familiar example is atmospheric ozone in Earth's stratosphere, which ultimately results from the photodissociation of molecular oxygen.  Yung \& DeMore (1999) provide a useful guide to these topics on many planets.
Photochemical products can themselves become important players in the atmospheric radiative transfer of giant planets, as discussed in the next section.

Photochemistry has long been expected (Marley 1998, Liang et al.\ 2004) to be important for hot Jupiter atmospheres and will likely be far more complex than in the solar system.  This is because molecular species that are condensed below the jovian clouds (e.g., $\rm H_2O,\ H_2S,\ NH_3$) and thus protected from photodissociation will be gaseous in such hot atmospheres.  Some of these species, such as  $\rm H_2S$, are easily photodissociated (Figure 5), and will likely produce new or unexpected species.  Sulfur and nitrogen compounds, in particular, may be important players in hot Jupiter photochemistry and perhaps haze production (Marley et al.\ 2007).   While the carbon photochemistry has been studied (Liang et al. 2004), preliminary work on  photochemistry in water-bearing $\rm H_2-He$ atmospheres suggests that compounds including $\rm CO_2$, HCN, and $\rm C_2H_6$ will be present well in excess of the abundance predicted by equilibrium chemistry (Troyer et al. 2007). Photochemical products may play a role in the formation of hot stratospheres (Marley 1998; Marley et al. 2007; Burrows et al. 2008a).  This area is certainly rich for further study.

Photochemistry can also be very important in terrestrial atmospheres as is the case with Venus and Titan.  An earthlike planet with a greater abundance of methane could well be enveloped in a photochemical haze and not appear anything like a `pale blue dot', even if there are indeed underlying oceans (Zahnle 2008).  Such obscuration by hazes is a major concern if planets are to be characterized solely by their colors (\S9).

\section{Temperature Inversions (Stratospheres)}

A gray atmosphere becomes asymptotically isothermal at small optical depths, reaching the skin temperature $T_0$ (\S2).   It is often the case that an atmosphere is optically thick to incident radiation at some wavelengths at low pressures where the atmosphere is simultaneously relatively transparent at infrared wavelengths.  In this case more incident energy may be absorbed than can be emitted by isothermal atmosphere with temperature $T_0$.  As a result the atmospheric layer with strong absorption must, in the absence of other energy transport mechanisms, heat up until the thermal emission from the layer equals the absorbed incident flux.  
 An inverted temperature structure with a warm, radiative upper atmosphere overlying a cooler tropopause, such as that shown for several planets' atmospheres in Figure 1, is known as a stratosphere.  Almost all solar system planets with an atmosphere exhibit a stratosphere.  In Earth's atmosphere ozone absorbs ultraviolet light, which warms the stratosphere to $270\, \rm K$,  about 50 Kelvin warmer than the temperature at the top of the troposphere.  
Solar system giant planet atmospheres are heated by UV absorption by a combination of methane and hydrocarbon photochemical products, including $\rm C_2H_2$ and $\rm C_2H_6$ and photochemically produced hazes (e.g., see the review for Neptune by Bishop et al.\ 1995).

The atmosphere of Jupiter provides a specific example.  Without an energy source the planet's middle atmosphere would be close to 104 K, (the skin temperature for Jupiter with $T_{\rm eff} = 124\,\rm K$), as seen above the tropopause in Figure 1.  In the region where most of the incident UV flux is absorbed (near 10 mbar) there is little overlap between a 100 K Planck function and the important thermal opacity sources, so little flux can be emitted.  Since the absorbed energy cannot be radiated away by a 100 K atmosphere, the atmosphere warms and the Planck function moves to shorter wavelengths.  Eventually the blue side of the Planck function overlaps the strong $\nu_4$ methane fundamental and the $\nu_9$ ethane bands at 7.7 and $12.2\,\rm \mu m$, allowing the atmosphere to radiatively cool, balancing the absorbed incident flux (Chamberlain \& Hunten 1987).  As in other solar system giant planet 
atmospheres, these strong mid-infrared bands of ethane and methane act as a thermostat, regulating the stratospheric temperatures. 

Likewise in exoplanet atmospheres a balance must be struck between the absorption of incident radiation and thermal emission.  For hot Jupiters which are so warm that even the most refractory Ti- and V-bearing compounds do not condense, TiO and VO gas may be exceptionally important absorbers (Hubeny et al.\ 2003; Burrows et al.\ 2007; Fortney et al.\ 2008). These gasses, while not abundant, have extraordinarily large absorption cross sections across the entire optical spectrum.  When present these molecules can absorb much of the remarkably high incident flux above $1\,\rm mbar$ where the atmosphere is optically thin in the thermal infrared.  The atmosphere therefore becomes very hot, perhaps as hot as $1900\,\rm K$ (Knutson et al. 2008), hot enough for emission by the near-infrared and optical bands of water, CO, and even TiO to balance this influx of energy (Fortney et al.\ 2008).  There is a hint from transit spectra of HD209458b that TiO and VO are indeed present in the atmosphere (Desert et al. 2008).

As in solar system atmospheres, photochemistry may also play an important role in exoplanet stratospheres.  Photochemical hazes or photochemical gaseous products that absorb well in the optical and UV could also provide prodigious energy sources for exoplanet stratospheric heating.  Photochemical pathways have not yet been fully explored for these planets (but see the work on the carbon chemistry by Liang et al. 2004 and speculation by Burrows et al. 2007) and the opacity spectra for many potential molecules are not well known.  This area remains ripe for further study.

\section{Albedos}

Albedo (from the Latin word for `white') is a measure of the reflectivity of an object.  In planetary science one encounters a variety of albedos (from single-scattering to geometric to Bond to spherical, to name a few) and care must be taken to carefully define the term in use.
From a planet-wide perspective the albedo of most importance is the Bond albedo, $A$, the ratio of incident energy reflected into all angles by a planet to the total incident energy.  The Bond albedo appears in the equation for the equilibrium temperature of a rapidly rotating planet with radius $R$ receiving an incident flux ${\cal F}$:
$$4\pi R^2 \sigma T_{\rm eq}^4 = (1-A) \pi R^2 {\cal F} \eqno (3)$$ 

The geometric albedo, $p_\lambda$, is defined as the ratio of a planet's reflectivity measured at zero phase angle (opposition) to that of a Lambert disk of the same radius.  Unlike the Bond albedo, the geometric albedo is a function of wavelength and, because it is measured at opposition (when the phase angle $\Phi=0$), does not require information on the dependence of scattering with phase to measure. For a perfectly reflecting Lambert sphere the geometric albedo is $2\over 3$; for a semi-infinite purely Rayleigh scattering atmosphere it is $3\over 4$.  Both such idealized, perfectly scattering objects would have a Bond albedo of 1, but the Rayleigh atmosphere sends more light directly back to the observer at zero phase angle and thus has a higher geometric albedo. Observed geometric albedo spectra for Uranus and Jupiter as well as a model spectrum for HD 209458b are shown in Figure 6.  Note that the hot Jupiter is quite dark beyond about $0.4\,\rm \mu m$ since  gaseous Na and K absorb most incident photons before they can be scattered back to space (see also Burrows et al. 2008b).  Uranus and Jupiter would likewise be quite dark in the red if not for their cloud layers that scatter red photons before they can be absorbed by methane (Figure 7; Marley et al. 1999).

\begin{figure}
\includegraphics[angle=90,scale=0.7]{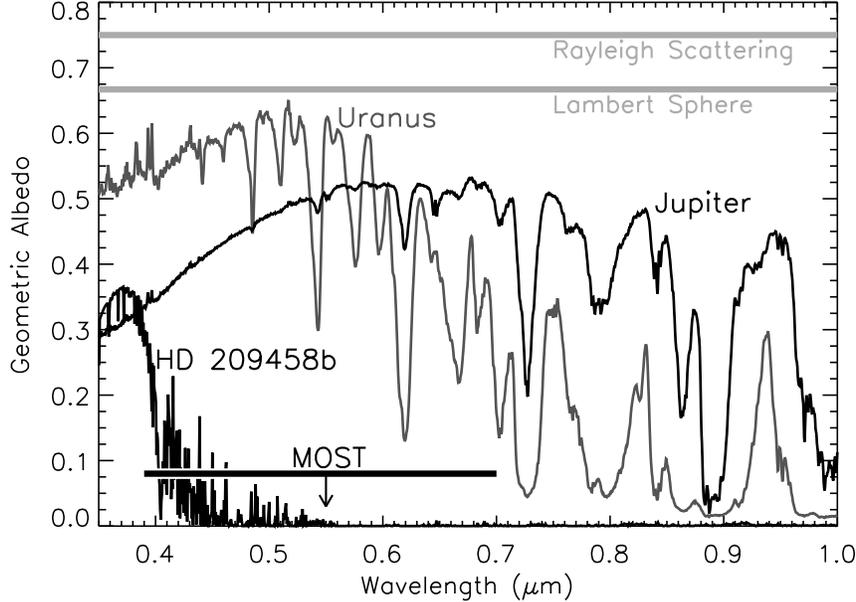}
\caption{Observed geometric albedo of Uranus and Jupiter (Karkoschka 1994), compared 
to one possible model of HD 209458b which ignores thermal emission, and the $1\sigma$ upper 
limit from MOST (Rowe et al. 2007). The geometric albedo of a deep Rayleigh scattering 
atmosphere and a Lambertian sphere are shown as well. Figure from Fortney et al. (in prep.). }
\end{figure}

The Bond albedo is related to the monochromatic geometric albedo $q_\lambda$ by
$$A=\int\limits_0^\infty p_\lambda q_\lambda f_\lambda d\lambda \Bigr / \int\limits_0^\infty f_\lambda d\lambda \eqno(2)$$
where $q_\lambda$ is the monochromatic phase integral and $f_\lambda$ is the incident monochromatic flux.  The phase integral is a measure of the angular distribution of scattered light.  For a Lambert sphere and a Rayleigh sphere $q_\lambda = {3\over2}$ and $4\over 3$ respectively.  
In the general case, however, a planet is neither a perfectly Lamberian nor a Rayleigh scatterer and the scattered light must be measured as a function of phase angle $\Phi$ and $\lambda$. This is why the Bond albedos for solar system giants could only be accurately determined after the {\it Voyager} spacecraft had measured their brightness at many different phase angles so that $q_\lambda$ could be measured (e.g. Pollack et al. 1986 for Uranus, see also the review by Conrath et al. 1989).  

Without a measurement of $q_\lambda$ it is common to assume that a planet scatters light isotropically following a Lambertian phase function.
The ratio $C$ of the brightness of a planet at a distance $a$ from its star as seen in reflected light at an arbitrary phase angle $\Phi$ to the brightness of the star can then be written as
$$C_\lambda(\Phi) = q_\lambda (R/a)^2 \Bigl [ {{\sin(\Phi) + (\pi-\Phi) \cos (\Phi)}\over \pi }\Bigr ]. \eqno(4)$$
At quadrature $\Phi = {\pi \over 2}$ and $C_\lambda = q_\lambda (R/a)^2/\pi$.  Note that for the giant planets in Figure 6 the contrast would be far more favorable in the visible than in the red or infrared bands.

\begin{figure}
\includegraphics[angle=90,scale=0.75]{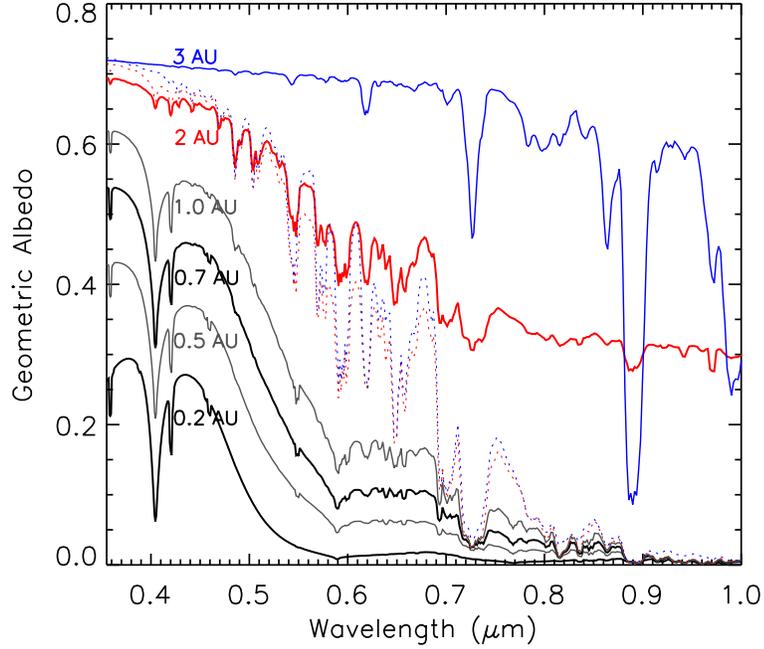}
\caption{Model geometric albedo spectra for a Jupiter-mass planet at various distances from a solar type star.  Dotted lines are models for the 3 and 2 AU planets but without cloud opacity.  While these spectra are cast as a function of orbital radius, the same sequence would result for a giant planet kept at a fixed distance and modeled with progressively younger ages, higher masses, or earlier stellar spectral types, all of which would produce warmer effective temperatures, all else being equal.  For example an $8\,\rm M_J$ planet at an age of about 1 Gyr would have a similar spectrum to the planet at 1.0 AU.  A $4\,\rm M_J$ at the same age would be similar to a cloudless planet (dotted line) at 2 AU.  Figure adapted from Fortney et al. (in prep.).}
\end{figure}

Because molecular bands tend to be stronger at longer wavelengths and because Rayleigh scattering is more efficient in the blue, most solar system giant planets have larger geometric albedos at shorter wavelengths than at longer wavelengths.  Given that the Bond albedo is weighted by the incident flux, which varies with stellar type, the same planet with the same geometric albedo spectrum will have a variety of different Bond albedos as the color of the incident flux is changed.  Around an early type star, which is brightest in the blue, a Jupiter-like planet might have a large Bond albedo of 0.45.  But around an M star the same planet's Bond albedo might be less than 0.1 (Marley et al. 1999).  For this reason it is usually best to consider the geometric albedo which does not overtly depend on the incident flux (other than through its dependence on the temperature profile).  

\section{Colors of Planets}

Early efforts to directly image extrasolar planets in reflected light, such as by small or intermediate-sized coronagraphs, will likely first produce images in a few broad passbands.  Given such limited available data it is worthwhile to consider what might be learned about the nature of the detected planets from such data.  

\begin{figure}
\includegraphics[angle=0,scale=0.7]{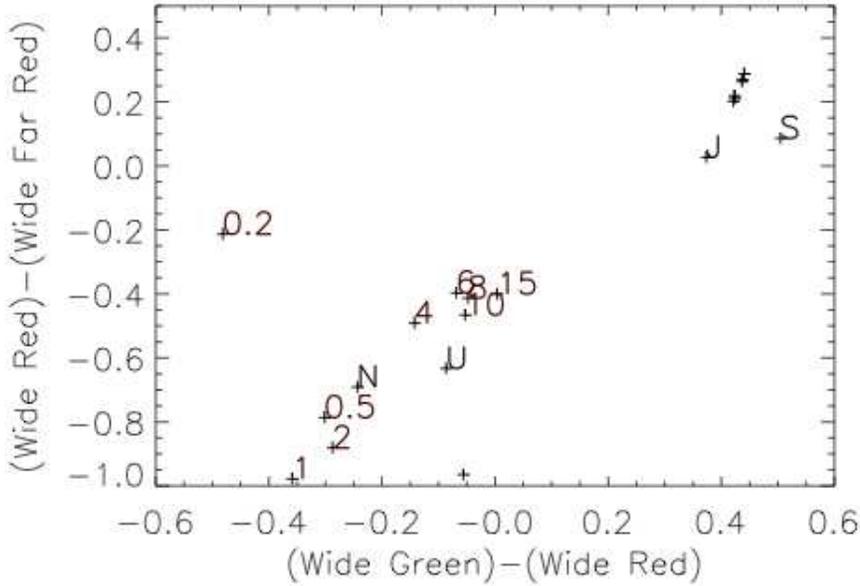}
\caption{Observed and model broad-band colors of giant planets.  {\it J, S, U,} \& {\it N} are the observed colors of the corresponding solar system giants.  Labeled crosses show model colors of giant planets at the given distances from a solar-type primary star and a simple cloud model.  Unlabeled crosses are Jupiter and Saturn models with a different set of assumptions about the cloud properties.  Depending on the assumed effective temperature and other model details (cloud structure, etc.), the color of a $1\,\rm M_J$ planet can vary widely.  Color alone is thus not a useful discriminant of mass.}
\end{figure}

Given an orbital separation from the primary star, a single photometric detection, combined with an assumed phase function and bounds placed on the geometric albedo, would allow a crude estimate of the planet's size.   Assuming an upper limit of $p<0.75$ (the pure Rayleigh scattering limit) and a lower limit $p>0.06$ (typical of low albedo asteroids (Dotto et al.\ 2002)), for example,  would result in an uncertainty in the radius inferred for a directly imaged planet of a factor of 3.5.  A bright planet with a radius slightly larger than Earth's could not be distinguished from a dark planet with Neptune's radius on the basis of brightness alone.  If the planet were also detected by other means, for example radial velocity or astrometric methods, then the known mass would discriminate between these two extremes.  Without such a detection, however, the nature of the planet would have to be discerned by spectroscopic or photometric methods.  Even low resolution spectroscopy likely will be beyond the reach of modest aperture space-based coronagraphic telescopes.  This means that planets will have to be characterized, at least initially, by their broadband colors.

Indeed based on our experience in the solar system, broadband colors of giant planets at first seem to be promising markers for discerning planet type (Figure 8 for the giants).  Uranus and Neptune are blue while Jupiter and Saturn are red.  Among the terrestrial planets the Moon, Mercury, and Mars are red while the  Earth is slightly blue.  It has been suggested (e.g., Traub 2003) that such color trends, when applied to exoplanets, may help identify the type of planets detected by direct imaging.  A planet might be imaged in a few broad spectral filters and characterized by comparison to solar system planetary colors.
There are substantial difficulties with this approach, however, since a single planet can have very different colors just depending on its temperature and the range of plausible colors is certainly much larger than that sampled by solar system planets.  The reflectivity of giant planets in the blue is strongly influenced by stratospheric hazes (e.g., see Baines \& Bergstralh (1986) for Uranus) while the brightness in the red depends upon cloud properties. As a result both photochemical hazes and clouds (Figure 7) can substantially alter the broadband color of a planet  and are difficult to model on an {\it a priori} basis.  
   Also there will certainly be exoplanets with characteristics quite different from solar system objects (warm Neptunes, super-Earths, water worlds, etc.) and their colors may be surprising.  

For example Figure 7 plots the geometric albedo spectra of a Jupiter mass planet at various distances from its primary star and thus warmer (see Marley et al.\ (1999), Sudarsky et al.\ (2003), and Burrows (2005) for more complete discussions).   An extrasolar planet in the same orbit could also be warmer than Jupiter if it were younger, more massive,\footnote{Since giants more massive than Jupiter take longer to cool, a four Jupiter mass planet that is 2 billion years old, would have $T_{\rm eff}\sim 400\,\rm K$ and exhibit a spectrum similar to one of the lower curves of Figure 7.} or orbiting a brighter and hotter star than our sun.  As we imagine warming Jupiter (Figure 3),  first the ammonia clouds would evaporate, allowing ammonia to be well mixed throughout the atmosphere and allowing us to see the water clouds, which underlie the ammonia clouds.  Since the water clouds are optically thick and good scatterers in the optical, such a planet would likely be much brighter and whiter than our current Jupiter.  As we continue to warm the planet the atmosphere would heat up and the cloud base would move progressively higher in the atmosphere.  Eventually, even the massive water clouds would evaporate, and suddenly we would have a clear atmosphere. Jupiter would then appear very dark blue (Figure 7), because red photons, which do not efficiently Rayleigh scatter, would burrow down into the atmosphere, never to return.  As the atmosphere continues to warm, alkali metals, which at low temperatures are found as chlorides such as KCl, would also evaporate, again altering the optical spectrum of the planet (Figures 6 \& 7) as they absorb strongly in the red (Burrows et al.\ 2000).  The important lesson from this thought experiment is that the same planet can have vastly different spectra--and by extension colors (Figure 8)--depending on its effective temperature. It would be a mistake to presume that one could recognize a Jupiter-mass planet simply by its broadband color.  In this example Jupiter varies from red to white to blue over the span of a few hundred Kelvin.  Spectra, such as those shown in Figure 7, along with inferences on age of the primary star (hence placing limits on the age of the planet) would be the best means of discriminating the nature of the planet.  

\section{Conclusions}
As we enter the age of the direct detection and characterization of extrasolar planets it is important not to overlook the lessons learned from the past half century of planetary exploration.  Solar system planets give ample demonstration of the importance of atmospheric dynamics, cloud and condensation processes, and photochemistry in controlling the face of planets.  The great diversity of atmospheres seen in our own solar system, from the hazy skies of Titan to the turbulent atmosphere of Jupiter to the blue vistas of Earth, are emblematic of the diversity of processes that can affect properties of planets.  Exoplanets will exhibit even larger ranges of properties and we should not be surprised by hazy yellow Earths, red, white, or blue Jupiters, or other unexpected worlds. Certainly extrasolar planet atmospheres will be influenced by some of the processes mentioned here.  (Other yet-to-be-discovered influences will undoubtedly be important as well.) Ultimately, however, such discoveries will extend the journey of planetary exploration begun by the {\it Mariners} and {\it Voyagers} of the last century out into the galaxy.

{\bf Acknowledgements:} The author thanks Jonathan Fortney and Kevin Zahnle for helpful comments on the manuscript and Thierry Montmerle and the entire Les Houches staff for organizing an exceptionally productive and informative program.



\begin{thebibliography}{99}

\bibitem[Baines 
\& Bergstralh(1986)]{1986Icar...65..406B} Baines, K.~H., \& Bergstralh, J.~T.\ 1986, Icarus, 65, 406 



\bibitem[Barman et al.(2005)]{2005ApJ...632.1132B} Barman, T.~S., 
Hauschildt, P.~H., \& Allard, F.\ 2005, Ap. J., 632, 1132 


\bibitem[Bishop et al.(1995)]{1995netr.conf..427B} Bishop, J., Atreya, 
S.~K., Romani, P.~N., Orton, G.~S., Sandel, B.~R., \& Yelle, R.~V.\ 1995, Neptune and Triton, 427 

\bibitem[Burrows(2005)]{2005Natur.433..261B} Burrows, A.\ 2005, Nature, 433, 
261 

\bibitem[Burrows et al.(2000)]{2000ApJ...531..438B} Burrows, A., Marley, 
M.~S., \& Sharp, C.~M.\ 2000, Ap. J., 531, 438 


\bibitem[Burrows et al.(2001)]{2001RvMP...73..719B} Burrows, A., Hubbard, 
W.~B., Lunine, J.~I., 
\& Liebert, J.\ 2001, Reviews of Modern Physics, 73, 719 

\bibitem[Burrows et al.(2006)]{2006ApJ...650.1140B} Burrows, A., Sudarsky, 
D., \& Hubeny, I.\ 2006, Ap. J., 650, 1140 


\bibitem[Burrows et al.(2007)]{2007ApJ...668L.171B} Burrows, A., Hubeny, 
I., Budaj, J., Knutson, H.~A., \& Charbonneau, D.\ 2007, Ap. J. Lett., 668, L171 

\bibitem[Burrows et al.(2008)]{2008ApJ...678.1436B} Burrows, A., Budaj, J., 
\& Hubeny, I.\ 2008a, Ap. J., 678, 1436 

\bibitem[Burrows et al.(2008)]{2008ApJ...682.1277B} Burrows, A., Ibgui, L., 
\& Hubeny, I.\ 2008b, Ap. J., 682, 1277 

\bibitem[1987]{cham1} Chamberlain, J. \& Hunten, D. 1987, {\it Theory of Planetary Atmospheres} (Orlando: Academic Press)

\bibitem[Charbonneau(2008)]{2008arXiv0808.3007C} Charbonneau, D.\ 2008, 
ArXiv e-prints, 808, arXiv:0808.3007 

\bibitem[Conrath et al.(1989)]{1989oeps.book..513C} Conrath, B.~J., Hanel, 
R.~A., 
\& Samuelson, R.~E.\ 1989, Origin and Evolution of Planetary and Satellite Atmospheres, 513 

\bibitem[Cowan 
\& Agol(2008)]{2008arXiv0806.4606C} Cowan, N.~B., \& Agol, E.\ 2008, ArXiv e-prints, 806, arXiv:0806.4606 



\bibitem[Deming(2008)]{2008arXiv0808.1289D} Deming, D.\ 2008, ArXiv e-prints, 808, arXiv:0808.1289 

\bibitem[Demory et al.(2007)]{2007A&A...475.1125D} Demory, B.-O., et al.\ 2007, Astron. \& Astrophys., 475, 1125 

\bibitem[Desert et al.(2008)]{2008arXiv0809.1865D} Desert, J.~-., 
Vidal-Madjar, A., Lecavelier des Etangs, A., Sing, D., Ehrenreich, D., 
Hebrard, G., \& Ferlet, R.\ 2008, ArXiv e-prints, 809, arXiv:0809.1865 

\bibitem[Dotto et 
al.(2002)]{2002A&A...393.1065D} Dotto, E., Barucci, M.~A., M{\"u}ller, T.~G., Brucato, J.~R., Fulchignoni, M., Mennella, V., \& Colangeli, L.\ 2002, A \& A, 393, 1065 



\bibitem[Fortney et al.(2007)]{2007ApJ...659.1661F} Fortney, J.~J., Marley, 
M.~S., \& Barnes, J.~W.\ 2007, Ap. J., 659, 1661 

\bibitem[Fortney et al.(2008)]{2008ApJ...678.1419F} Fortney, J.~J., 
Lodders, K., Marley, M.~S., \& Freedman, R.~S.\ 2008, Ap. J., 678, 1419 


\bibitem[Helling et al.(2008)]{2008arXiv0809.3657H} Helling, C., et al.\ 
2008, MNRAS in press, arXiv:0809.3657 



\bibitem[Hubeny et al.(2003)]{2003ApJ...594.1011H} Hubeny, I., Burrows, A., 
\& Sudarsky, D.\ 2003, Ap. J., 594, 1011 

\bibitem[Ingersoll \& Porco(1978)]{1978Icar...35...27I} Ingersoll, A.~P., \& Porco, C.~C.\ 1978, Icarus, 35, 27 

\bibitem[Karkoschka(1994)]{1994Icar..111..174K} Karkoschka, E.\ 1994, Icarus, 111, 174 


\bibitem[Knutson et al.(2007)]{2007Natur.447..183K} Knutson, H.~A., et al.\ 
2007, Nature, 447, 183 



\bibitem[Knutson et al.(2008)]{2008ApJ...673..526K} Knutson, H.~A., 
Charbonneau, D., Allen, L.~E., Burrows, A., 
\& Megeath, S.~T.\ 2008, Ap. J., 673, 526 

\bibitem[Liang et al.(2004)]{2004ApJ...605L..61L} Liang, M.-C., Seager, S., 
Parkinson, C.~D., Lee, A.~Y.-T., \& Yung, Y.~L.\ 2004, Ap. J. Lett., 605, L61 

\bibitem[Lodders(2004)]{2004..Science} Lodders, K.\ 2004, Science, 303, 323 

\bibitem[Marley(1998)]{1998ASPC..134..383M} Marley, M.~S.\ 1998, Brown 
Dwarfs and Extrasolar Planets, 134, 383 

\bibitem[Marley et al.(2007)]{2007prpl.conf..733M} Marley, M.~S., Fortney, 
J., Seager, S., \& Barman, T.\ 2007, Protostars and Planets V, 733 

\bibitem[Niemann et al.(1998)]{1998JGR...10322831N} Niemann, H.~B., et al.\ 
1998, J.\ Geophys.\ Res., 103, 22831 

\bibitem[Pollack et al.(1986)]{1986Icar...65..442P} Pollack, J.~B., Rages, 
K., Baines, K.~H., Bergstralh, J.~T., Wenkert, D., 
\& Danielson, G.~E.\ 1986, Icarus, 65, 442 



\bibitem[Rowe et al.(2007)]{2007arXiv0711.4111R} Rowe, J.~F., et al.\ 2007, 
ArXiv e-prints, 711, arXiv:0711.4111 

\bibitem[Seager et al.(2008)]{2008arXiv0808.1913S} Seager, S., Deming, D., 
\& Valenti, J.~A.\ 2008, ArXiv e-prints, 808, arXiv:0808.1913 



\bibitem[Showman et al.(2007)]{2007arXiv0710.2930S} Showman, A.~P., Menou, 
K., \& Y-K.~Cho, J.\ 2007, ArXiv e-prints, 710, arXiv:0710.2930 

\bibitem[Showman et al.(2008)]{2008arXiv0809.2089S} Showman, A.~P., 
Fortney, J.~J., Lian, Y., Marley, M.~S., Freedman, R.~S., Knutson, H.~A., 
\& Charbonneau, D.\ 2008, ArXiv e-prints, 809, arXiv:0809.2089 

\bibitem[Sudarsky et al.(2003)]{2003ApJ...588.1121S} Sudarsky, D., Burrows, 
A., \& Hubeny, I.\ 2003, Ap. J., 588, 1121 

\bibitem[Traub(2003)]{2003ASPC..294..595T} Traub, W.~A.\ 2003, Scientific 
Frontiers in Research on Extrasolar Planets, 294, 595 

\bibitem[Troyer et al.(2007)]{2007DPS....39.2203T} Troyer, J., Moses, 
J.~I., Fegley, B., Lodders, K., Marley, M.~S., 
\& Fortney, J.~J.\ 2007, Bulletin of the American Astronomical Society, 38, 450 



\bibitem[Wallace(1976)]{1976jupi.conf..284W} Wallace, L.\ 1976, Jupiter, 
284 


\bibitem[Yung \& Demore(1999)]{1999ppa..conf.....Y} Yung, Y.~L., \& Demore, W.~B.\ 1999, Photochemistry of planetary atmospheres / Yuk L.~Yung, William B.~DeMore.~New York : Oxford University Press, 1999.

\bibitem[Zahnle(2008)]{2008Nature454} Zahnle, K.\ 2008,  Nature, 454, 41 


\end{thebibliography}
\end{document}